\documentstyle[preprint,prb,aps]{revtex}
\begin{document}
% \draft command makes pacs numbers print
\draft
\title{$CVD$ routes to $MgB_2$ conductors}
% repeat the \author\address pair as needed
\author{D. K. Finnemore, W. E. Straszheim, S. L. Bud'ko, P. C. Canfield, 
and N. E. Anderson, Jr }
\address{Ames Laboratory, U.S. Department of Energy and Department of Physics and Astronomy\\
Iowa State University, Ames, Iowa 50011}
\author{Raymond J. Suplinskas}
\address{Specialty Materials, Inc.\\
1449 Middlesex Street, Lowell MA 01851}

\date{\today}
\maketitle
\begin{abstract}
% insert abstract here
Processing methods are described for the development of magnesium diboride wire 
using the chemical vapor deposition ($CVD$) to produce long lengths of suitably doped starting 
boron fiber.  It is found that titanium can be co-deposited with the boron to make long lengths 
of doped fiber that contain both $TiB$ and $TiB_2$.   When this fiber is reacted in $Mg$ 
vapor to transform boron into $MgB_2$, the resulting conductor has a superconducting critical 
current density of about $5\times 10^6~A/cm^2$ at $5~K$ and self field.  The critical 
current density at $25~K$ and $1$ Tesla is $10,000~A/cm^2$. Using optical methods to 
define grain boundaries and energy dispersive X-rays to determine $Ti$ and $Mg$ 
concentration, these samples show a fine dispersion of $Ti$ through out the grains and 
no conspicuous precipitation of $TiB_2$ on the $MgB_2$ grain boundaries. This is to be 
contrasted with the precipitation of $TiB_2$ on $MgB_2$ grain boundaries observed for 
samples prepared by solid state reaction of $Ti$, $Mg$, and $B$ powders.  Introducing 
$Ti$ impurities into the $B$ during the $CVD$ deposition of the $B$ gives a  
distribution of $TiB_2$ in the resulting $MgB_2$ different from solid state reaction of powders.

\end{abstract}
% insert suggested PACS numbers in braces on next line
\pacs{74.25.Bt, 74.25.Fy, 74.25.Ha}

% body of paper here

\section{Introduction}

The discovery\cite {1} of superconductivity in $MgB_2$ at $39.2~K$ opens some 
very interesting opportunities  for practical applications of this material as a conductor 
in power applications.\cite {2}  Typically, the operating temperature of these 
conductors would be at about half the 
superconducting transition temperature, $T_c$, or about $20~K$ where  closed cycle refrigerators are quite efficient.  Early studies of  the temperature and magnetic field dependence of  the superconducting 
critical current density, $J_c$,\cite {3} indicated that grain boundaries in this material would carry very 
substantial supercurrents, $~5\times 10^5~A/cm^2$,  in conductors  prepared 
in a variety of ways.\cite{4,5}  In addition, a very early attempt to explore the use of powder-in-tube methods by Jin and co-workers\cite {6} showed that practical conductors are 
feasible.

On the fundamental side, the measurement of an isotope\cite {7} effect with $\alpha =0.26$ where 
$T_c\sim M^{-\alpha }$ and $M$ is the isotopic mass of the boron atom indicated that 
phonons play a major role in the superconducting interaction in $MgB_2$.  For clean 
samples, resistivity 
measurements at $40~K$ indicated that $\rho \sim 1.0\mu ohm-cm$ which, combined with the 
calculated Fermi velocity, \cite {8} $v_F\sim 4.7\times 10^6~m/s$, gives a mean-free-path, 
$\ell \sim 60 nm$.  From the measured upper critical field, $H_{c2}~16~T$.\cite {9}, the 
coherence distance of a clean sample can be determined from  $H_{c2}=\Phi _o/[2\pi \xi_o ^2]$ 
to be $\xi _o\sim 5~nm$.  Here, $\Phi _o$ is the flux quantum.  With $\ell $ substantially greater than $\xi _o$, these materials are in the clean limit.  Several groups\cite {4,5} found $J_c$ 
values in the range of $500,000~A/cm^2$ in self field at $5~K$, but these values fell below 
$1,000~A/cm^2$ at $1~T$ at $5~K$. Broadly speaking, the irreversibility field, $H_{irr}$ was  
roughly half of $H_{c2}$.

The primary drawback for this material appeared to be the relatively low values of the 
upper critical field, $H_{c2}=16~T$,\cite {9} and the resulting rapid drop in $J_c$ with field.  
 The most straight forward way to raise $H_{c2}$ is to shorten 
the mean-free-path, $\ell $, which in turn shortens the coherence distance, 
$\xi =\sqrt {\xi _o\ell }$.  Thus $H_{c2}=\Phi _o/[2\pi \xi ^2]$ would rise with decreasing $\ell $. 
The introduction of point defects is an effective way to raise $H_{c2}$.  In pure $MgB_2$, 
the volume pinning force seems to decrease as $[1-H/H_{irr}]^2$ indicating that grain boundary 
pinning dominates.\cite {4} Hence, reducing the grain size should help.  Alternatively, the 
strength of the pinning and $J_c$, can be increased by introducing precipitates.  
Typical dimensions might be an array of pinning sites with a characteristic dimension of the coherence distance, $1-5~nm$, and a spacing between pinning centers about $5$ times larger.

Doping of $MgB_2$ has proven to be difficult because relatively few impurities go into the 
lattice in solid solution.  Early reports by Slusky and co-workers\cite {10} show that $Al$ added by reacting powders in a solid state reaction substitute for $Mg$ in the $MgB_2$ lattice.  There is a smooth 
change in lattice parameter and smooth decrease in $T_c$ for $Mg_{1-y}Al_yB_2$ for $y$ 
out to $0.10$ and then a two phase region is entered.  At $y=0.1$, $T_c$ is still about $33~K$ 
and thus still relatively high.  

For $Ti$ doping by solid state reaction of powders, it appears that relatively little $Ti$ is incorporated in the the $MgB_2$ grains.  Work by Zhao and coworkers\cite {11}showed that 
$Ti$ appears as very thin layers of $TiB_2$ precipitates  
on the $MgB_2$ grain boundaries.  Samples were made via solid state reaction by grinding the 
$Ti$, $Mg$, and $B$ powders, pressing pellets, and heat treating.
Microstructural analysis seemed to indicate that $TiB_2$ 
may act as a sintering agent leading to higher density, very fine $MgB_2$ grain-size of about $10~nm$, and excellent grain-to-grain coupling.   Some high resolution transmission electron microscopy evidence points to the thickness of some of these grain boundary layers being as low as one $TiB_2$ unit cell.\cite {11}  
Because the $Ti$ does not enter the unit cell, there is relatively little diminution 
in $T_c$ and very substantial increases in the critical current density, $J_c$.  Optimum doping 
was in the range of  a $Mg/Ti$ ratio of about $0.1$.  One of the more important outcomes of 
this work was the observation that the $Ti$ seems to refine the grain structure to a grain size of 
about $10~nm$.

For carbon doping, rather large amounts of carbon can replace boron in the structure.  
 Takenobu and co-workers\cite {12} found that carbon substitutes for boron in 
$MgB_{2-x}C_x$ out to $x\sim 0.06$ as indicated by a smooth decrease in a-axis lattice 
parameter from $0.3085~nm$ at $x=0$ to $0.3070~nm$ at $x=0.06$ while $T_c$ drops from 
$38.2~K$ to about $34.8~K$.  There was relatively little change in the c-axis dimension.  
 Here, samples were made by mixing powders of the elements and 
firing at $900^{\circ }C$ for $2~h$.  
At higher $x$-values, they found that 
the (110) X-ray peak of $MgB_2$ 
broadens indicating two phases.\cite {12} .  Bharathi and co-workers,\cite {13} 
by contrast, found that reacting under 
high pressure $Ar$, permitted the extention of carbon content to $x=0.5$.  They find a relatively smooth depression 
of $T_c$ from $39~K$ at $x=0$ to about $26~K$ at $x=0.5$.  Using a somewhat different
procedure, we have  found that 
reacting $B_4C$ with $Mg$ vapor at $950^{\circ }C$ for $2~h$  gives a material with  
$T_c\sim 17~K$.  In other work, the change in lattice constant and $T_c$ with carbon content also has been 
reported to be somewhat less than these values.\cite {14}

For Si doping, a solid state reaction of $Mg$ and amorphous $B$ powder 
with nano-size $Si$ particles show substantial 
$Mg_2Si$ lines in the X-ray data and relatively little diminution in the transition temperature to  about $36~K$.\cite {15} This seems to imply that very little Si seems to enter the $MgB_2$ structure.  In an alternate preparation method, Cooley and co-workers\cite {16} have shown $Mg$ can 
be reacted with $SiB_6$ at temperatures as low as $500^ {\circ }C$ to give $Mg_2Si$ 
precipitates in a $MgB_2$ matrix with $T_c=37~K$.  These preliminary experiments seem to 
indicate that the enhancements in flux pinning arise from $Mg_2Si$ precipitates rather than 
$Si$ substitution into the $MgB_2$ lattice.

The purpose of this review is to summarize progress in  attempts to make $MgB_2$ conductors 
using a CVD method for depositing the starting boron material.  
There is a well established 
commercial scale method to make boron 
fiber in kilometer lengths for use in high strength composites.  The primary goal  of this work is to find ways to make doped boron fiber that will permit the 
preparation of long lengths of high performance $MgB_2$ 
conductor.  Tasks include 1) the development of processing methods to convert pure 
boron fiber to $MgB_2$,  2) the development of methods to dope 
the starting boron 
fiber with point defects 
that will raise the upper critical field, $H_{c2}$,  3) the development of  boron fibers that will produce nano-scale precipitates in the $MgB_2$, and 4) the development of methods 
to handle the brittle character of the superconductor.   

Chemical vapor deposition of the starting boron is a particularly attractive route 
to materials preparation because the impurity atoms can possibly be introduced with 
atomic scale dispersion.  Initial experiments have focussed on  titanium-doping to produce 
small  $TiB_2$ precipitate pinning centers with the next step to be the introduction of  carbon  point defects to shorten the electronic mean free path.

In a typical 
deposition process for boron fibers, a conducting filament such as tungsten or a carbon 
coated commercial silicon carbide fiber is introduced 
into a long glass chamber through a $Hg$ seal at a rate of a few $cm/s$.   A flowing stream of $BCl_3$ and hydrogen gas move through the full length of the chamber.  The fiber is heated electrically to temperatures in the $1100-1300^{\circ }C$ range, 
and the boron is deposited in the hot zone.  If a suitable carrier gas can be mixed with this 
gas stream, then impurity atoms such as 
$Ti$ or $C$ can be co-deposited with the boron to form filaments that can subsequently 
converted to impurity doped-$MgB_2$.  Critical current densities were measured via magnetization as reported earlier.\cite {3}

\section {Conversion of pure $B$-fiber to magnesium diboride}

Soon after the discovery of superconductivity\cite {1} in $MgB_2$, small wire segments of 
commercial boron fiber were converted to $MgB_2$.\cite {17} 
These wire segments were only a few 
$cm$ long and rather fragile, but they showed exceptionally low resistivity of $0.4~\mu ohm-cm$ 
at $40~K$ and a  sharp superconducting transition at $39.2~K$.  In a  subsequent series of 
upper critical field measurements,\cite {9,17} $H_{c2}vsT$ 
curves indicated an extrapolation to $H_{c2}(T=0)$ of about $16~T$. Critical current densities, $J_c$ for a $100~\mu m$ diameter pure boron fiber reacted in $Mg$ vapor at $900^{\circ }C$ for $2~h$ are shown for both 
$5~K$ and $25~K$ in Fig. 1.  At $5~K$, $J_c\sim 5\times 10^5~A/cm^2$ in self field, and 
$J_c$ is about $1000~A/cm^2$ at $2.5~T$.  At $25~K$, $J_c\sim 2\times 10^5A/cm^2$ 
in self field, and  $J_c$ is about $1000~A/cm^2$ at $1.4~T$.  Also shown on Fig. 1 are data 
for a sample made by depositing $25~\mu m$ of pure boron on a carbon substrate.  The $5~K$ 
data are the open circles and the $25~K$ data are the solid triangles.  Within the reproducibility 
of the experiments, the $100~\mu m$ pure boron fibers give the same $J_c$ values
as samples made from $25~\mu m$ of pure B on a C substrate within the accuracy of the 
measurement.  For all samples, the reaction is carried out at a single temperature, and the time 
can then be adjusted to optimize $J_c$ if that is desired.  

A series of measurements were undertaken to determine the lowest temperatures where 
these boron fibers would transform to $MgB_2$ at a high rate.  A sample reacted at $800^{\circ }C$ for 
$2~h$, as shown in Fig. 2, shows a $Mg$ region only part way across the boron fiber in 
the relatively light area near the edge of the filament.  A more detailed analysis shows that this 
is a  uniform wall of a high boron compound, $MgB_x$, 
progressing across the fiber.  Energy dispersive spectra in the scanning electron microscope 
indicate that this phase has $x$ near $6$ or $7$.  Because this phase is not the study of 
this work, further analysis was not undertaken.  For this time and temperature, essentially no $MgB_4$ or 
$MgB_2$ forms.  The white core of the filament is $W_2B_5$.  

By contrast, if a similar pure boron fiber is heated in $Mg$ vapor at 
$875^{\circ }C$ for $2 h$, the fiber transforms to $MgB_2$ nearly everywhere as shown in 
Fig. 3.  The darker gray areas are small regions of $MgB_4$ as shown by the EDS line scan 
along the line shown, where the circle with a cross is the starting point of the scan.  The large dip 
in $Mg$ count rate is the $W_2B_5$ core and the $30$ percent dip in count rate to the left of the core illustrates the decrease in $Mg$ density going from $MgB_2$ to $MgB_4$.   Also visible in 
the micrograph are several faults along the edge of the fiber and a large void near the right end 
of the portion of the $W_2B_5$ core.  The filament is not completely flat, so that the $W_2B_5$ 
core does not show along the entire length of the filament for this particular depth of polish.  
These flaws are typical in $MgB_2$ 
filaments prepared from commercial pure $B$ fibers.  In Fig. 4, an electron micrograph is shown in end view for the long time limit where a $100~\mu m$ diameter fiber that has been reacted in $Mg$ vapor at $950^{\circ }C$ for $78~h$.  The white core of $W_2B_5$ in the center is 
measured here to be $16.7~\mu m$ in diameter and the starting $100~\mu m$ boron fiber has grown to about $135~\mu m$ diameter as $Mg$ is added.  There is a small patch of $MgB_4$ 
in the upper left quadrant, and there the void pattern is very typical of all the samples grown.

\section {Titanium doping}

In the early experiments with $CVD$ doping of $B$ with $Ti$, the material was deposited on either a $W$ substrate or a $C$ coated substrate, and 
a preliminary report of this work has been presented.\cite {18}  
Most of  the experiments reported here, however,  
were done by simultaneously depositing $B$ and a few percent $Ti$ on a continuous 
$75~\mu m$ diameter silicon carbide filament coated with several micrometers of glassy 
carbon (commercially available as SCS-9A).   Deposition was accomplished in a reactor 
similar to that used for commercial boron filament production.\cite {19} The substrate 
filament was drawn through the reactor at $100~mm/s$; the filament in the reactor 
was resistively heated to peak temperatures of approximately $1100^{\circ }C$.  The 
reactant gas was predominantly a stoichiometric mixture of hydrogen and boron trichloride 
at atmospheric pressure.  Titanium dopant was added by bubbling the hydrogen component 
through titanium tetrachloride held at $0^{\circ }C$.  The resulting partial pressure of 
titanium tetrachloride was approximately $2\times 10^{-3}~bar$.  The doped 
boron-coated substrate was spooled continuously upon emerging from the reactor.
An X-ray powder diffraction study of the resulting fibers before $Mg$ reaction indicates comparable amounts of $TiB$ and $TiB_2$ as well as a small amount of $B_{10}C$.  Subsequent reaction with $Mg$ vapor 
produced $MgB_2$ filaments with precipited $TiB_2$.  The intensity of the X-ray lines indicates 
less $TiB$ in the reacted fiber than in the unreacted fiber.  

A series of samples were prepared in which $B$ and $Ti$ were co-deposited on the  
carbon surface with thicknesses of  $4~\mu m$, $6~\mu m$ and $10~\mu m$.  After 
reaction in $Mg$ vapor at both $950^{\circ }C$ for $2~h$ and at $1100^{\circ }C$ for 
$30~min$, the conversion of $B$ to $MgB_2$ was complete.\cite {18}  
An extensive study of the $Ti$
distribution was undertaken with $EDS$ in a $SEM$ where the beam size was approximately $1$ 
micrometer.  Generally, the $Ti/Mg$ ratio was in the neighborhood of $9\%$  in the bulk of 
the material.  This is illustrated in Fig. 5 where a sample with $6~\mu m$ of $B\sim 9\%Ti$ 
is deposited on a carbon coated substrate.  There are two regions of the sample, however, 
that are consistently high $Ti$.  This is near the $C$ substrate 
and near the outer surface.  Reaction kinetics indicate that the deposition of $Ti$ compared to $B$ is higher
at lower temperatures.  Hence, as the fiber moves through the deposition zone, the $Ti$ deposition 
will be faster than that of boron during the initial warm-up on entrance to the hottest regions 
and during the cool-down as the fiber exits the hottest regions. Another specimen illustrated in 
Fig. 6 shows a $4~\mu m$ thick layer of $B\sim 9\% Ti$.  Here, there 
are two rather distinct regions of $Ti$ content.  In most of the boron layer, the $Ti$ 
content is about $9\% $, but sometimes there are 
bright lacy regions where the 
$Ti$ content is about $30\%$.  By studying polished regions of the $MgB_2$ with polarized 
light, the grain structure is readily apparent.  If a particular region of the sample is 
studied both with polarized light and with $SEM$ with $EDS$ area maps, 
the $Ti$ is found to be  
spread throughout the grains. The grain boundaries are not seen to be high in $Ti$ on these  
area maps.  This result differs from the beautiful decoration of $MgB_2$ grain boundaries with 
$TiB_2$ seen for samples made from powders in solid state reactions.\cite  {11}  Titanium 
doping using $CVD$ methods gives a dispersion of $TiB_2$ in $MgB_2$ that is different from 
that seen for solid state reaction of powders.

\section {Critical current densities}

The addition of $Ti$ greatly enhances the $J_c$ values if the level of $Ti$ doping is properly 
adjusted.  As reported previously, anything approaching a $Ti/Mg$ ratio of 1 is far too much 
$Ti$.\cite {18}  The volume fraction of superconductor is reduced by half, and the value of $T_c$ drops to 
about $20~K$.  In accord with earlier findings,\cite {9} a $Ti/Mg$ ratio in the neighborhood 
of $0.09$ greatly enhances both $J_c$ and $H_{irr}$ while depressing $T_c$ only a few 
degrees.  At $5~K$ and self field, $J_c$ increases 
from about $4\times 10^5~A/cm^2$ for pure $MgB_2$ to about 
$2\times 10^6~A/cm^2$ for the $9\%~Ti$ sample.  With increasing magnetic 
field, $J_c$ crosses $10,000~A/cm^2$ at $5~T$ with $9\%~Ti$ as compared to $1.4~T$ 
for the pure $MgB_2$ sample as shown by the solid triangles in Fig. 7.  At $25~K$, the 
$9\% Ti$ sample crosses $10,000~A/cm^2$ at about $1.3~T$ compared with the 
pure $MgB_2$ sample which crosses $10,000~A/cm^2$ at $0.9~T$.  

In addition to simply adding $Ti$ to the boron fibers, 
there is a great deal of optimization that can yet be done. 
When the starting $B_{0.91}Ti_{0.09}$ fibers are prepared, there is a large amount 
of both $TiB$ and $TiB_2$ in the boron.  During the reaction to form $MgB_2$, some of 
the $TiB$ is converted to $TiB_2$, and there may be some migration of $C$ from the 
substrate into the sample.  Much more needs to be learned about both of these processes.  An 
optimization of $J_c$ at $25~K$ with  $B_{0.91}Ti_{0.09}$\cite {18} shows that 
a reaction time of $30~min$ is best for a reaction temperature of $1100^{\circ }C$ and 
a reaction time of $2~h$ is best for a reaction temperature of $950^{\circ }C$.
An optimization for $J_c$ values at $5~K$, shown in Fig. 8, indicates that $30~min$ at 
$1100^{\circ }C$ gives the highest $J_c$ values.

\section {Thickness dependence}

For pure $B$ fibers deposited on a $15~\mu m$ $W$ core, a series of $B$ fibers having 
diameters ranging from $100~\mu m$ to $300~\mu m$ were fully reacted in $Mg$ vapor at 
$950^{\circ }C$ for times ranging from $2~h$ to $48~h$.  In all cases, the reaction to form 
$MgB_2$ increases the fiber diameter by about $35\% $.  At $5~K$ and self field, the 
$J_c$ values range from $700,000~A/cm^2$ to $300,000~A/cm^2$ and have a magnetic field 
dependence similar to those shown in Fig. 1.  At $25~K$, the data of Fig. 9 indicate that 
$J_c(H)$ nearly is independent of diameter.  To the accuracy and reproducibility of the measurement, 
all of these fiber diameters have the same $J_c~vs~H$ plots.

For the  $B9\% Ti$ deposited on a carbon coated substrate, samples having a thickness of 
$4~\mu m$, $6~\mu m$, and $10~\mu m$ have been studied.   
We expected a difference in $J_c$ among these samples 
caused by different amounts of $C$ impurity 
or caused by a different distribution of $TiB$ or $TiB_2$ during the $Ti$ and $B$ deposition on 
the $C$ substrate. These are relatively small effects as shown on Fig. 10.  These data were 
taken for samples reacted at $1100^{\circ }C$ for $15~min$. A reaction time less than 
the $30~min$ for optimum $J_c$ was intended to limit possible diffusion of $C$ from the 
substrate.  In broad terms, the $5~K$ 
$J_c$ data are about $3\times 10^6~A/cm^2$ at low field and the $J_c$ data cross the $10,000~A/cm^2$ 
line at about $4~T$.  There is not a large thickness dependence of $J_c$, but at high fields, 
the thicker samples seem to show a higher $J_c$ values. At low fields, the thinner samples seem 
to show the higher $J_c$ values. 
Similar data have been reported elsewhere.\cite {18} 
There is lots of room for improvement 
by optimizing the $Ti$ content, the reaction temperatures, and the reaction times.

\section {Conclusions}

Co-deposition of  boron along with the desired impurities offers a distinctive method of doping.  When these doped fibers are converted to 
$MgB_2$ by heating in $Mg$ vapor, the resulting $TiB_2$ is distributed relatively uniformly within the grains.  This distribution is quite different from 
$MgB_2$ samples prepared by solid state diffusion of mixed powders.  Critical current densities of $2$ to $5$ million $A/cm^2$ are 
obtained at $5~K$ and self field.  Critical current densities of $10,000~A/cm^2$ are obtained 
at $25~K$ and $1.4~T$.
 
A number of important variables are still not well understood. 
Improvement is needed in the control of the distribution of $TiB_2$, 
and a more 
thorough study of  carbon doping  needs to be undertaken so that a quantitative 
contrast can be developed 
between the effects of $C$ substitution in the lattice with the effects of 
$TiB_2$ precipitates.  Although there is a fairly wide sweet spot for the reaction of doped $B$ fibers 
to form doped $MgB_2$, there is lots of room for improvement.   
The performance of the conductor can yet be optimized with respect to 
$Ti$ content and carbon content as well as the   
time and temperature of $Mg$ reaction.

\section{Acknowledgments}
Work is supported by the U.S. Department of Energy, Basic Energy Sciences, 
Office of Science, through the Ames Laboratory under Contract No. W-7405-Eng-82.

\vfil\eject

\begin{figure}
\caption{Comparison of $J_c$ values for  pure $MgB_2$ made from $100~\mu m$ boron 
fibers with a $W_2B_5$ core and $25~\mu m$ of pure boron on a carbon substrate 
at both $5~K$ and $25~K$. } 
\end{figure}

\begin{figure}
\caption{Micrograph and EDS line scan showing the initial diffusion of $Mg$ into a $100~\mu m$ diameter boron fiber.  The white core is $W_2B_5$} 
\end{figure}

\begin{figure}
\caption{Typical section of a $MgB_2$ filament that is fully reacted.  The white area in the center
is a portion of the $W_2B_5$ core.  The darker gray areas are $MgB_4$ as indicated by 
the $30\%$ drop in $EDS$ scan intensity.} 
\end{figure}

\begin{figure}
\caption{Typical end view of a $MgB_2$ filament prepared from a $100~\mu m$ diameter
pure $B$ fiber.  The dark gray area in the upper left quadrant is $MgB_4$ and void area is  
typical.} 
\end{figure}

\begin{figure}
\caption{A section of $4~\mu m$ thick layer of $B9\%Ti$ deposited on a $C$ substrate and 
reacted at $1100^{\circ }C$ for $15~min$.  The light and dark regions reflect a change in 
$Ti/Mg$ ratio.  High $Ti$ near the $MgB_2$ inner and outer surfaces is typical.} 
\end{figure}

\begin{figure}
\caption{A section of a $4\mu m$ thick layer of $B9\%Ti$ deposited on a $C$ substrate
in which the $Ti$ is high in the outer half and about $50\%$ as high in the inner half.}
 \end{figure}

\begin{figure}
\caption{Comparison of the $J_c$ data for a sample made from $25~\mu m$ of pure 
boron on a carbon substrate with a sample made from $10~\mu m$ of $B_{0.91}Ti_{0.09}$ 
on a carbon substrate at both $5~K$ and $25~K$.  Both $J_c$ and $H_{c2}$ are increased substantially by adding a few percent $Ti$.} 
\end{figure}

\begin{figure}
\caption{Optimization of $J_c$ vs time at $1100^{\circ }C$ and $9\% Ti$.} 
\end{figure}

\begin{figure}
\caption{$J_c$ data for various wire diameters for pure $B$ fiber at $25~K$.} 
\end{figure}

\begin{figure}
\caption{$J_c$ data for various thicknesses of $B_{0.91}C_{0.09}$ at $5~K$.} 
\end{figure}

\vfil\eject


\begin{references}  

\bibitem{1} J. Nagamatsu, N. Nakagawa, T. Muranaka, Y Zenitani, and J. Akimitsu, 
Nature {\bf 410} (2001) 63.
\bibitem{2} Paul M. Grant MRS Symposium Proceedings, {\bf 689}
(2002) 3.
\bibitem{3} D. K. Finnemore, J. E. Ostenson, S. L. Bud'ko, G. Lapertot, P. C. Canfield,
Phys. Rev. B {\bf 86} (2001) 2420.
\bibitem{4} D. C. Larbalestier, M. Rikel, L. D. Cooley, A. A. Polyanskii, J. Y. Jinang, S. Putnaik, 
X. Y. Cai, D. M. Feldmann, A. Gurevich, A. A. Squitier, M. T. Naus, C. B. Eom, E. E. 
Hellstrom, R. J. Cava, K. A. Regan, N. Rogado, M. A. Hayward, T. He, J. S. Slusky, K. Iaumaru, M. Haas, Nature {\bf 410} (2001) 186. 
\bibitem{5} Y. Bugoslavsky, G. K. Perkins, X. Qi, L. F. Cohen, and A. D. Caplin,
Nature {\bf 410} (2001) 563.
\bibitem{6} S. Jin, H. Mavoori, and R. B. vanDover, Nature, {\bf 411} (2001) 563.
\bibitem{7} S. L. Bud'ko, G. Lapertot, C. Petrovic, C. E. Cunningham, N. E. Anderson, 
and P. C. Canfield, Phys. Rev. Lett. {\bf 86}, (2001) 1877. 
\bibitem{8}J. Kortus, I. I. Mazin, K. D. Belashchenko, V. P. Antropov, and L. L. Boyer, 
Phys. Rev. Lett. {\bf 86} (2001) 4656.
\bibitem {9} S. L. Bud'ko, C. Petrovic, G. Lapertot, C. E. Cunningham, P. C. Canfield, 
M - H, Jung, and A. H. Lacerda, Phys. Rev. B, {\bf 63} (2001) 220503(R). 
\bibitem {10} J. S. Slusky, N. Rogado, M. A. Hayward, P. Khalifah, T. He, K. Inumaru, 
S. Loureire, M. K. Haas, H. W. Zandbergen, and R. J. Cava, Nature, {\bf 410} 2001 343.
\bibitem{11} Y. Zhao, Y. Feng, C. H. Cheng, L. Zhou, Y. Wu, T. Machi, Y Fudamoto, 
N. Koshizuka, and M. Murakami, Appl. Phys. Lett. {\bf 79}, 1154 (2001); Y. Zhao, D. X. 
Huang, Y. Feng, C. H. Cheng, T. Machi, N. Koshizuka, and M. Murakami, Appl. Phys. Lett. 
{\bf 80} (2002) 1640. 
\bibitem {12} T. Takenobu, T. Ito, D. H. Chi, K. Prassides, and Y. Iwasa, Phys. Rev. 
{\bf 64} (2001) 134513.
\bibitem {13} A. Bharathi, S. K. Balaselvi, S. Kalavathi, G. L. N. Reddy, V. S. Sastry, Y. Haribaran, and T. S. Radfakrishnan, 
Physica C {\bf 370} (2002) 211.
\bibitem {14} M. Paranthaman, J. R. Thompson, and D. K. Christen, Physica C {\bf 355} 
(2001) 1.

\bibitem {15} X. L. Wang, S. H. Zhou, M. J. Qin, P. R. Monroe, S. Soltanian, H. K. Liu, and 
S. X. Dou, LANL Cond. Mat./0208349 
\bibitem {16} L. D. Cooley, V. Braccini, J. Waters, P. Hellenbrand, B. Senkowicz, J. Y. Jiang, 
P. J. Lee, E. E. Hellstrom, and D. C. Larbalestier, (private communication)
\bibitem {17} P. C. Canfield, D. K. Finnemore, S. L. Bud'ko, J. E. Ostenson, G. Lapertot, 
C. E. Cunningham, and C. Petrovic, Phys. Rev. Lett. {\bf 86} (2001) 2423.
\bibitem {18} N. E. Anderson, W. E. Straszheim, S. L. Bud'ko, P. C. Canfield, and 
D. K. Finnemore,  Physica C,  Submitted .
\bibitem {19} R. J. Suplinskas and J. V. Marzik, "Boron and Silicon Carbide 
Filaments", in \underline {Handbook of Reinforcements for Plastics}, J. V. Milewski and 
H. S. Katz, ed., Van Nostrand Reinhold, New York, (1987).






\end{references}
\end{document}